\documentclass[11pt,twoside]{article}
\usepackage{psfig,amsfonts,amssymb,amsbsy}
\usepackage[mathscr]{eucal}

\newcommand{\Lapl}{{\boldsymbol{\nabla}^2}}
\newcommand{\Gkst}{G_{\!\mbox{\tiny $K$}}(s,t)}
\newcommand{\Gkonest}{G_{\!\mbox{\tiny $K$}}^1(s,t)}
\newcommand{\Gktwost}{G_{\!\mbox{\tiny $K$}}^2(s,t)}
\newcommand{\Gki}[2]{G_{\!\mbox{\tiny $K_{#2}$}}(#1,t)}
\newcommand{\Gkoneat}[2]{G_{\!\mbox{\tiny $K$}}^1(#1,#2)}
\newcommand{\Gktwoat}[1]{G_{\!\mbox{\tiny $K$}}^2(#1,t)}
\newcommand{\Sphi}{{S_\phi(\vec k,t)}}
\newcommand{\Xiktwo}{{\Xi_{\!\mbox{\tiny $K$}}^2(x)}}
\newcommand{\rescS}{{\mathscr S}}
\newcommand{\vK}{{\vec{K}}}
\newcommand{\sump}[2]{\sum_{#1}^{#2}\!\!\raisebox{0.4ex}{$'$}\;\;}

\newcommand{\de}{\mbox{d}}
\newcommand{\ddt}[1]{{\left({\partial{#1}\over\partial t}\right)_{\!\!\alpha}}}
\newcommand{\dddaa}[1]{{\partial^2{#1} \over \partial\alpha^2}}
\newcommand{\dddss}[1]{{\partial^2{#1} \over \partial s^2}}

\newcommand{\eq}[1]{eq.~(\ref{eq:#1})}
\newcommand{\noeq}[1]{(\ref{eq:#1})}

\newcommand{\Equ}[1]{Equation~(\ref{eq:#1})}
\newcommand{\fig}[1]{fig.~\ref{fig:#1}}

\newcommand{\dontput}[1]{}

\begin{document}

\centerline{Appeared in Euro. Phys. J. {\bf B9}, 719-724 (1999).}

\noindent{\LARGE\bf Curvature autocorrelations in domain\\growth
dynamics}\\

\noindent\hspace{0.8cm}\parbox{13.5cm}{
  {\Large Oliver L.~Sch\"onborn and Rashmi C.~Desai ($^1$)}\\

  ($^1$) Department of Physics, University of Toronto, Toronto, Ontario
  M5S 1A7 CANADA; e-mail: desai@physics.utoronto.ca\\

  (Received April 21, 1998)\\

  PACS.05.20.-y -- Statistical mechanics\\
  PACS.82.20.Mj -- Nonequilibrium kinetics\\
  PACS.71.45.G  -- Correlation functions (collective effects)\\
  PACS.02.60.Cb -- Numerical simulation; solution of equations\\

  {\bf Abstract.}  ---We show how the interface curvature autocorrelation
  function (ICAF) and associated structure factor (ICSF), of relevance in
  non-equilibrium pattern-formation problems where sharp interfaces are
  present, provide new and interesting information on domain structure, as yet
  not visible via the order-parameter structure factor (OPSF).  This is done by
  discussing numerical simulations of model A (non-conserved relaxational
  phase-ordering kinetics) in two-dimensional systems.  The ICAF is Gaussian
  over short distances and exhibits dynamical scaling and $t^{1/2}$ power-law
  growth. We use it to show what the typical length-scale in the model A 
  dynamics corresponds to physically and how it can be obtained uniquely,
  rather than simply within a multiplicative constant.  Experimental methods to
  measure the ICAF and/or ICSF are still needed at this point.}

\vspace{1.5cm}




\section{Introduction}

Dynamics of pattern-formation in non-equilibrium systems is a very
challenging problem \cite{guntonSS83}.  It is ubiquitous in nature and
its understanding is of interest and importance in physical and biological
sciences. It is customary to use light or particle scattering experiments on 
such systems to measure the spatial or temporal correlations, yielding a 
scattering function or {\em structure factor}.  In systems which can be 
described by an order-parameter, the order-parameter structure factor (OPSF) has
traditionally been the quantity of interest, as it is easily measurable
experimentally and is well-defined mathematically, making comparison
between theory and experiment possible.  In principle, the OPSF
contains much of the structural information of the system's state at a given
time.
However, simulations of the well-known \cite{guntonSS83} model A system,
described below, show that some fundamental configurational information is not
readily (if at all) visible with the OPSF.  Experimental methods for
measuring the ICAF or ICSF do not seem to exist at present.  Results
discussed here show a clear need for such experimental measurement
methods.

Hohenberg and Halperin in the late 70's proposed a classification for
several types of pattern-forming dynamics for which a field-theoretic
description existed \cite{hohenbergH77}.  One of the classes, labeled
model A, is that of dissipative dynamics for a single uncoupled non-conserved
order-parameter $\phi$.  The order-parameter could be for instance the
local magnetization in an Ising ferromagnet, i.e.~an idealized
ferromagnet in which the magnetization of any molecule can take only two
values, either $+1$ (up) or $-1$ (down).  Model A is described by a
Time-Dependent Ginzburg-Landau (TDGL) equation relating the temporal and
spatial variations of the order-parameter of the system:
\begin{equation}
  M^{-1}{\partial\phi\over\partial t} = \phi - \phi^3 + \xi^2\Lapl \phi
  \label{eq:modelA}
\end{equation}
where $M$ and $\xi$ are positive phenomenological constants determining
the time-scale and interface length-scale of the dynamics, respectively.
Some phenomenological parameters have been scaled out.  In this equation,
there is no constraint on the average order-parameter per unit area as a
function of time.  This differs from the well known model B which describes
spinodal decomposition in binary mixtures \cite{guntonSS83}, and for which the
order parameter is conserved.

Under appropriate conditions, such as a critical quench, model A
dynamics is characterized by the formation of convoluted, interpenetrating 
domains of two phases.  The domains are separated from one another by sharp
interfaces, i.e.~sharp on the length-scale of the domains but smooth and of
finite-width on the length-scale of the molecules of the system \cite{oonoP87}.
Once the interfaces have formed, the system enters the so-called scaling
(i.e.~late stage) regime, where the dynamics is strongly non-linear.  Note that
the width of interfaces remains approximately constant throughout the
late-stage regime, and is roughly $5\xi$.  In this regime, the system seeks to
decrease the amount of interface via interface motion.

There are several interesting aspects to the dynamics during the late-stage
regime, but for the purposes of this article we recall only two of them.
First, experiments and numerical simulations observe self-similar dynamics,
whereby the system state at a given time is statistically the same as that
at a later time, if space is properly rescaled\dontput{ using the concepts
of dynamic renormalisation group}.  The minimal condition for
this is that the dynamics must have the same time-dependence on all
length-scales, so that all dynamical lengths can be expressed in terms of
one arbitrarily chosen reference length-scale $L(t)$ which encompasses the
unique time-dependence. The second important characteristic of model A dynamics
is that $L(t)\sim t^{1/2}$.

Mathematically, the OPSF is written
\begin{equation}
  \Sphi \equiv
    {1\over A}\left<\hat\phi(\vec k,t)\hat\phi(-\vec k,t)\right>
\end{equation}
where $A$ is the system area, $\hat\phi(\vec k,t)$ is the Fourier transform
of the order-parameter $\phi(\vec x, t)$ at time $t$, and the angle brackets
denote, as usual, an ensemble average over all possible initial system
configurations.  $S_\phi(\vec k,t)$ gives the
statistical intensity of each mode of the order parameter field $\phi$,
i.e.~the importance of domains whose coarse size is $2\pi/|\vec k|$.
$\Sphi$ itself is the Fourier transform of the two-point
equal-time order-parameter correlation function
\begin{equation}
G_\phi(\vec x,t) \equiv \left< \phi(\vec x_0+\vec x,t)
                               \phi(\vec x_0,t)  \right>,
\end{equation}
which is a measure of the spatial correlations in the order-parameter field
at a given time.  The main characteristics of $S_\phi(\vec k,t)$ for model A
are a maximum at $\vec k=0$, and for large $\vec k, S_\phi\sim |\vec
k|^{-(d+1)}$, where $d\equiv $~dimensionality of space.  This power-law tail
is known as Porod's law and is a direct consequence, as shown by
Porod \cite{porod82}, of the sharpness of interfaces.
The $\vec k=0$ dominant mode indicates that the probability of finding larger
domains increases with the (coarse) size of the domain, and hence 
there is no $typical$ coarse size for the domains.  
This is in stark contrast to
model B dynamics, where $S_\phi(\vec k,t)$ shows a peak at non-zero $\vec k$.

The most important consequence of self-similar dynamics is that
$S_\phi(\vec k,t)$ taken at different times during the scaling regime can be 
rescaled in amplitude and $\vec k$, using $L(t)$ described earlier, to fall on 
one, universal and time-independent curve $\rescS_\phi(y)$, with
$y\equiv|\vec k|L(t)$ (hence the denomination of ``scaling'' regime).  
$L(t)$ can be computed, for instance, from any moment of $S_\phi(\vec k,t)$.
Regardless of how it is computed, it will always have  the same time
dependence, {\em within} a multiplicative factor.  The unique time
dependence has lead to the conclusion that a typical, or dominant, length scale
exists in model A dynamics.  But the arbitrariness of the multiplicative factor
has eluded attempts at computing or even identifying its physical origin
from the order parameter correlation functions.  Here, we solve this problem 
by focusing on the dynamics at the domain interfaces.

Indeed, the late time dynamics of model A is known to be dominated by the
motion of interfaces which decouple from the interior of the domains.  The
interface dynamics is curvature driven according to the Allen-Cahn result
\cite{cahn}:  the interface velocity at a given interfacial point is
proportional to its curvature at that point; the proportionality
coefficient is $M~\xi^2$ where the phenomenological constants $M$ and $\xi$ are
respectively the mobility and interfacial width (assumed time independent at 
late times). While much has been written about this interface dynamics, it 
pertains only to the behaviour normal to the interface. In contrast, 
correlations along the interfaces as they evolve have to our knowledge not been
explored.  We report here for the first time some new results for the space and
time dependent curvature-curvature  correlations along the interface.



\section{Results}

\begin{figure}[t]
  \centerline{\psfig{figure=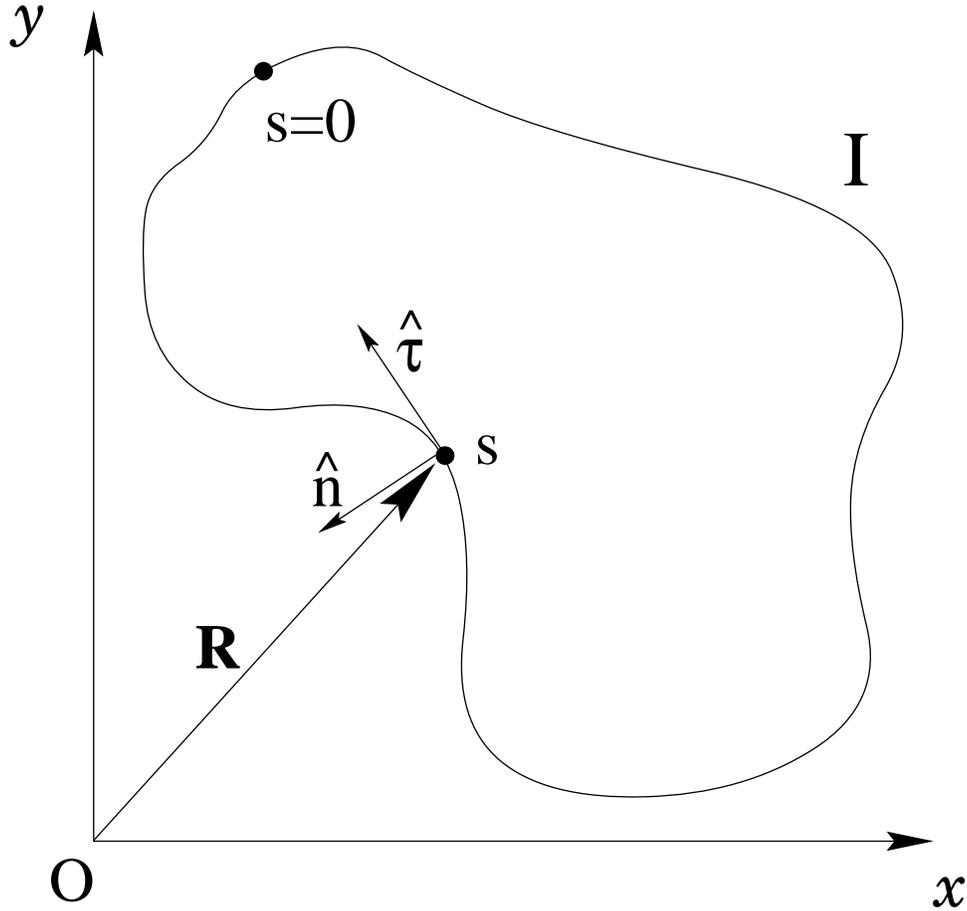,width=5in,angle=270}}
  \caption{Parametric representation of an interface in the plane, as a
  function of arclength $s$ along the interface.}
  \label{fig:interf}
\end{figure}

An interface {\bf{I}} in the plane can be represented parametrically by a set of
vectors
$\vec R(s) = (x(s), y(s))$ defined relative to an arbitrary origin, as
in
\fig{interf}.  The parameter $s$ is the arclength along the interface.
In this notation, the curvature at point $s$ is $\vec K \equiv \dddss{\vec
R}$.

The generic definition of the ICAF is
\begin{equation}
  \Gkst \equiv \left< \vK(0,t)\cdot \vK(s,t) \right>
  \label{eq:Gkst}
\end{equation}
where $\vK(s,t)$ is the curvature of an interface as a function of arclength
position $s$ along the interface and time $t$.  Numerically there are at
least two ways of computing $\Gkst$, assuming the interfaces are discretized
homogeneously with a mesh of size $\Delta s$.  They both use \eq{Gkst} for
interface $i$, which takes the form
\begin{equation}
  \Gki{s}{i} = {1\over n_i(t)}
               \sum_{j=1}^{n_i(t)} \vK_i(s_j,t) \cdot\vK_i(s_j+s,t).
\end{equation}
where $s$ is assumed an integer multiple of $\Delta s$, the subscript $i$
refers to interface $i$, $n_i(t)$ is the number of points on the interface,
and $K_i(s_j,t)$ is the curvature of the interface at some point
$s_j=j\Delta s$, at time $t$.

The first method is a simple weighted average of all $\Gki{s}{i}$, giving
more importance to longer interfaces.  We denote it by $\Gkonest$,
\begin{equation}
  \Gkonest \equiv {1\over \sum_{i=1}^{N_I(t)}{}' n_i(t)}
                  \sump{i=1}{N_I(t)} n_i(t) \Gki{s}{i}
  \label{eq:Gkone}
\end{equation}
where the prime superscript in the sum indicates that only interfaces longer
than a length of $2s$ are used and $N_I(t)$ is the number of interfaces used.

The second definition, which uses the same notation as $\Gkonest$, is
\begin{equation}
  \Gktwost \equiv {1\over \sum_{i=1}^{N_I(t)}{}' n_i(t)}
        \sump{i=1}{N_I(t)} n_i(t) \left( {\Gki{s}{i}\over\Gki{0}{i}}
\right),
  \label{eq:Gktwo}
\end{equation}
i.e.~$\Gktwost$ is a weighted average of the $normalized$ ICAF for each
individual interface.  Hence $\Gktwost\leq1$ for all $s$, with exact
equality at $s=0$, and can be interpreted as the average {\em relative}
value of curvature a distance $s$ on either side of a point where the
curvature is $\vK$.  \Equ{Gkone} turned out to be more suitable for
analytical calculations, while \eq{Gktwo} gives smaller statistical error
numerically, but both measure the same correlation of local interfacial
curvature.

\begin{figure}[t]
  \centerline{\psfig{figure=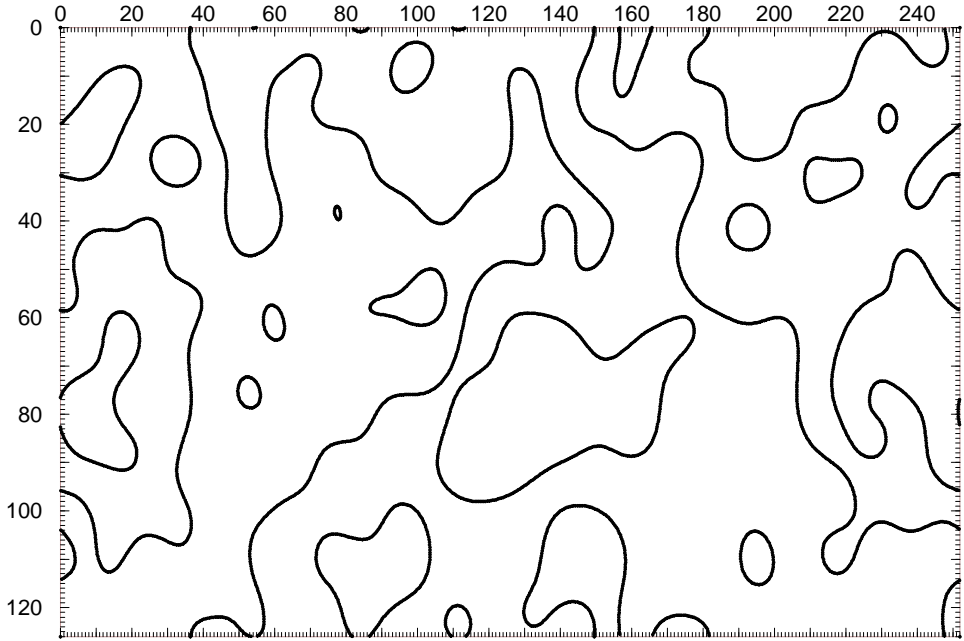,width=5in,angle=180}}
  \caption{Domain interfaces in a typical two-dimensional model A
system.}
  \label{fig:typical}
\end{figure}
Numerical simulations of model A were done for flat Euclidean systems of
sizes $100\times100$ and $200\times200$, with periodic boundary conditions.
Forty random order-parameter configurations were generated and evolved by
integrating \eq{modelA} using a standard Forward-in-time/Centered-in-Space
Euler integration scheme \cite{NRinC92}, from $t=0^+$ to $t=1000$.  The
system mesh size $\Delta x$ used was 1, and the time step $\Delta t=0.03$.
The results were checked to be independent of system-size.  Average
computation time required was 8 hours for the 40 runs on an HP735.  A
typical interface configuration is shown in \fig{typical}.

Computing $\vK$ is very difficult if the bulk description (\eq{modelA}) is
used, as it requires extracting interfaces from the order-parameter
configurations by systematically scanning these and finding all interfaces,
splining them for smoothness and finally computing $\vK$ at regular
intervals along an interface.  An interface description, which evolves the
interfaces directly via an interface equation \cite{schoenbornD98}, allows for
direct computation of $\vK$ and therefore $\Gkst$, while the runtime can be
decreased by a factor of 10 for flat systems and 50 to 100 for the curved
(i.e.~non-Euclidean) model A \cite{schoenbornD98}. Such a discretized interface
description produced results {\em closer} to the analytic predictions
than did the discretized bulk description ({\em cf.} Discussion section),
but both yielded otherwise identical results.

\begin{figure}[t]
  \centerline{\psfig{figure=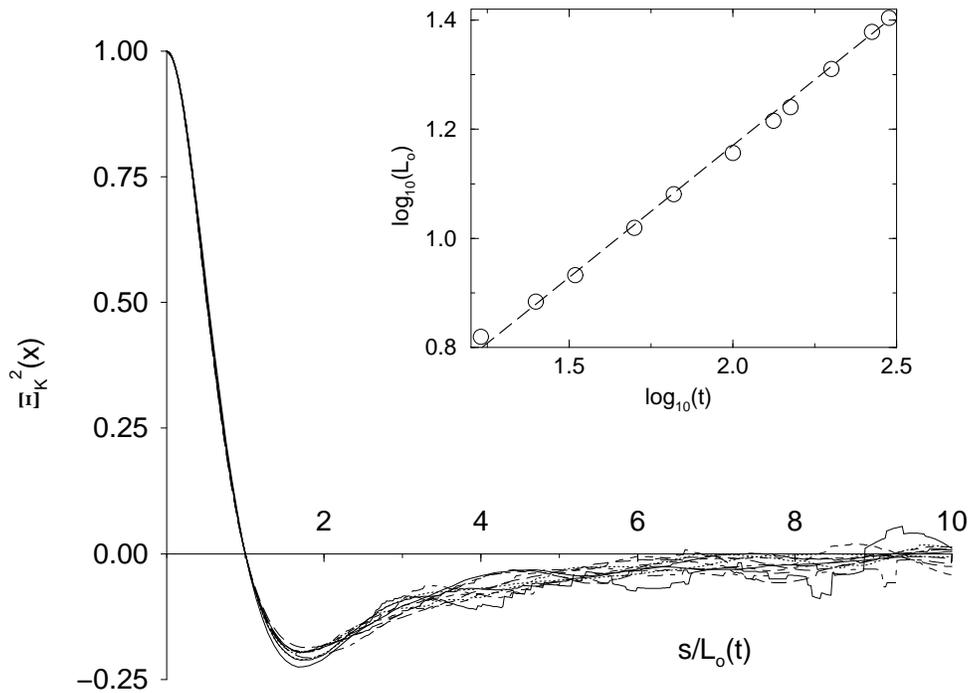,width=5in,angle=270}}
  \caption{$\Xiktwo$, from bulk configurations at various times.  $L_o(t)$
was defined as the first zero of $\Gktwost$.}
  \label{fig:Gk4stIFallresc}
\end{figure}

The ICAF was computed at several different times during the scaling regime,
$t=17$ to $t=300$.  The rescaled ICAF, denoted
\begin{equation}
  \Xiktwo \equiv \Gktwoat{L_o(t)^{-1}s}, \;\;\; x\equiv L_o(t)^{-1}s,
  \label{eq:xikx}
\end{equation}
is shown in \fig{Gk4stIFallresc}, where the scaling length-scale $L_o(t)$
was arbitrarily defined as the first zero of $\Gktwost$.
The vertical error bars, not shown for clarity, are 0 at the origin and
increase roughly linearly to 0.01 in the vicinity of the minimum, then
further increase to 0.02 at $s=200$.  The error was computed by making an
analogy between the curvature $\vK$ and magnetization $\vec m$ of
one-dimensional Ising magnets of different lengths \cite{mkrumbhaarB73}.  The
error for distance $s$ is then the weighted average of the deviation of each
magnet's value of $\langle m(0)m(s)\rangle$ from the value of $\langle
m(0)m(s)\rangle$ for the ensemble of magnets.  $\Gktwost$ corresponds to
$\langle m(0)m(s)\rangle$.  This was deemed the most reasonable method of
error calculation, given the values of $\vK$ along an interface, and
therefore the statistical error in products of $\vK$, are correlated.



\section{Discussion}

The salient features of the curves in \fig{Gk4stIFallresc} are the nice
superposition of the $\Xiktwo$ (within error bars), the power law for
$L_o(t)$, the negative autocorrelation for $s>L_o(t)$, and finally the
Gaussian form for distances much smaller than $L_o(t)$.
The perfect dynamical scaling indicates that the ICAF correctly captures
this very important characteristic of model A dynamics, and that the number
of runs and system size used give an accurate and reliable measure of
$\Gktwost$.  The power law in $L_o(t)$ is found through linear regression to
be $0.45\pm0.02$, very close to the theoretical value \cite{bray94} of 1/2
(the interface description gives a power-law even closer to the analytic
prediciton of 1/2, namely $0.48\pm0.01$, as it is less sensitive to
discretization effects).  $\Gktwost$ therefore also captures to a high
accuracy the well-known power-law behavior characteristic of model A dynamics.

Over short distances (up to roughly $L_o(t)/2$), $\Gkst$ can be checked to be
Gaussian.  A Gaussian form for a correlation function was shown by Porod to
be a characteristic of fluctuating systems without sharp
variations \cite{porod82}.  A plot of $K(s)$ (not shown) indeed looks very
much like a snapshot of a one-dimensional fluctuating ``membrane'', smooth
and without any sharp variations ({\em cf.}  \fig{typical}).  A first
attempt at obtaining an approximate analytical expression for $\Gkonest$ is
exposed here, based on a method developed by T.M. Rogers in his Ph.D.~thesis
\cite{rogers}, in the context of model B, but never tested numerically.

Consider the model A curvature equations \cite{schoenbornD98} in a flat system:
\begin{eqnarray}
  \ddt K        & = & \dddss K + K^3 \label{eq:dKdtRogers}\\
  \ddt{\sqrt g} & = & -\sqrt g K^2   \label{eq:dgdtRogers}
\end{eqnarray}
where $s$ is the arclength along an interface, $\alpha$ is a parameterization 
of the interface in which every point has constant
coordinate $\alpha$ (exploiting the fact that the interface moves
perpendicularly to itself \cite{schoenbornD98}), and $g$ is the metric
on the interface, relating the elements of length in both gauges ($s$ and
$\alpha$):
\begin{equation}
  \de s = \sqrt g \;\de\alpha.
  \label{eq:dsvsda}
\end{equation}

Now let us make the following mean-field approximation:
\begin{eqnarray}
  \ddt{\sqrt g} & \simeq & -\sqrt g h(t) \label{eq:dgdtapprox}\\
           h(t) & \equiv & {1\over L}\int\!\!K^2\; \de s
\end{eqnarray}
where $L$ is the total length of the interface.  
This approximation becomes exact
for circular domains.  Furthermore, neglect the cubic term of
\eq{dKdtRogers}.  This term is dominant for circular domains.  For
convoluted domains, numerical testing indicates $K^3$ is comparable to the
diffusion term for as many as half the interface points.  Therefore, the two
approximations work in opposite directions, one becoming exact for circular
domains, the other getting better for convoluted domains.  On short
length-scales, however, convoluted domains are locally circular, but the
$K^3$ term should be negligible since convoluted domains see their curvature
decrease rather than increase.  \Equ{dKdtRogers} then becomes
\begin{equation}
  \ddt K  =  {1\over g(t)}\dddaa K .
  \label{eq:dKdt2}
\end{equation}
Going to Fourier space and making use of a change of variable for time,
$g(t)\de t' = \de t$, the integration can now be performed, and the curvature
structure factor $\chi_q$ obtained:
\begin{equation}
  \chi_q(t') \equiv {1\over N} <K_q(t') K_{-q}(t')>
               = {<K_q(0)^2>\over N} e^{-2q^2t'}
\end{equation}
where $N$ is the number of points on the interface and $q$ is the wavenumber
in the reciprocal space of $\alpha$.  Assuming all $K_q$ have equal
amplitude at $t=0$, a backwards Fourier transform yields
\begin{equation}
  \Gkoneat{\alpha}{t'} = \sqrt{\pi\over8t'} e^{-\alpha^2/8t'} .
  \label{eq:Gkapprox}
\end{equation}
Now $t'$ must be found as a function of $t$.  This can be done by noting
that $h(t)$ is equal to $\Gkoneat{0}{t'(t)}$, i.e.~$h(t)=\sqrt{\pi\over8t'}$. 
Also,
\[
  t'(t) = \int_0^t\!\! {\de t\over g(t)} .
\]
The equation for the metric is therefore
\begin{equation}
  \ddt{g} = -2 g \sqrt{\pi\over8}
    \left( \int_0^t\!\! {\de t\over g(t)} \right)^{-1/2} .
  \label{eq:dgdt3}
\end{equation}
This has $g(t)=t^{-1}/\pi$ as solution, so that $t'(t)=\pi t^2/2$.  From
\eq{dsvsda}, $\alpha=s/\sqrt g=s\sqrt{\pi t}$.  Substituting in \eq{Gkapprox},
\begin{equation}
  \Gkonest = {1\over2t} e^{-s^2/4t},
  \label{eq:Gkapproxfinal}
\end{equation}
which is Gaussian, has time-dependencies consistent with power-law growth in
model A dynamics (the amplitude of $\Gkonest$ has units length squared), and
dynamically scales, but does not capture the negative autocorrelations at
longer distances.  Quantitatively, numerical simulations find the amplitude
of the Gaussian to go as $[2(1.1\pm0.1)t]^{-1}$, in very good agreement with
\eq{Gkapproxfinal}, whereas the width is smaller than the prediction
\noeq{Gkapproxfinal} by about a factor of two.

Of course, the two approximations that allowed for the calculation of
$\Gkst$ are reasonable only on short length-scales, so it is not surprising
that \eq{Gkapproxfinal} does not capture the dominant wavelength of
undulations of the interfaces.  Also, all $K_q(0)$ were assumed equal.
This is obviously wrong, since even at the earliest times the curvature
structure factor (the Fourier transform of the $\Gkst$) shows a well-defined
peak at a non-zero $q$ mode.  If the correct $\chi_q(0)$ is used, then the
analytical $\Gkst$ will have a dip at least at early times.  Therefore the
strongest approximation may yet lie in the $\chi_q(0)$ rather than the
mean-field and linearization approximations, though this seems unlikely.

\begin{figure}[t]
    \centerline{\psfig{figure=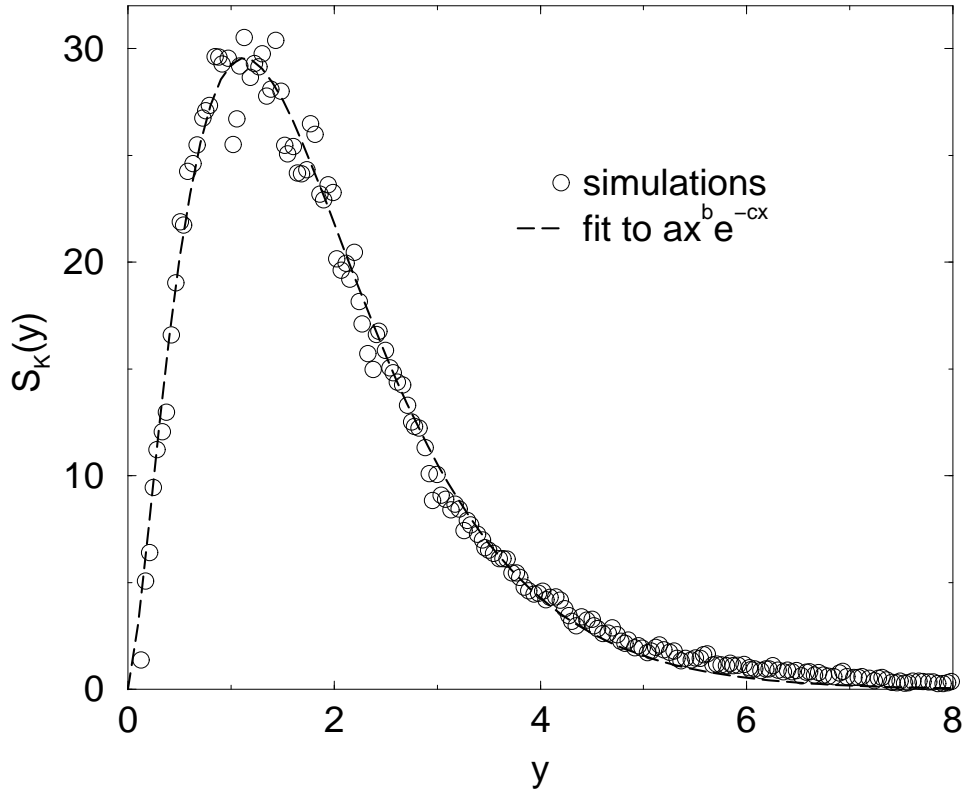,width=5in,angle=270}}
    \caption{The rescaled curvature structure factors $\rescS_K(y)$, as a
             function of $y\equiv2\pi L(t)/s$, for the same times as in
             \fig{Gk4stIFallresc} (joined into one curve for clarity). Fit
             gives $a=110, b=3/2$ and $c=4/3$ with 3\% error.}
    \label{fig:fftGk2}
\end{figure}
The most interesting feature of the ICAF is undoubtedly the relatively large
negative autocorrelation apparent at distances $s > L_o(t)$.  
Hence the rescaled 
ICSF $\rescS_K(y)$ (Fourier transform of \eq{xikx}) shows a
well-defined peak at a non-zero value of wavevector $y\equiv2\pi L_o(t)/s$, as
seen in \fig{fftGk2}.  The functional form can be fitted very nicely to
$ay^be^{-cy}$ for $0<y\lesssim 5$, with $a=110\pm3, b=3/2\pm0.01$ and
$c=4/3\pm0.04$, though no theoretical justification for this is known to
us at present.  The maximum of this function occurs at $y=b/c=9/8\pm0.04$,
which is extremely close to 1.  The null value of $\rescS_K(0)$ stems from the 
null area under the $\Gkonest$ curves, itself a direct consequence of
\begin{equation}
  \int \!\!\vK_i \;\mbox{d} s = 0
  \label{conservK}
\end{equation}
for any interface $i$ (when the system has periodic boundary conditions),
a constraint akin to a conservation law.  The peak in $\rescS_K(y)$ indicates 
that there is a {\em dominant wavelength} for undulations of the interfaces.  
This dominant spatial mode is a feature which remains as yet either hidden in 
or inaccessible via the order-parameter structure factor.  To our knowledge, it
is the first time that it is clearly identified.  The closeness of $y$ to 1 
also indicates that the first zero of $\Gkonest$ corrresponds to the dominant 
spatial oscillation mode for curvature undulations, 
i.e.~the typical arclength distance from a point on the interface at which the 
curvature changes sign.

\begin{figure}[t]
    \centerline{\psfig{figure=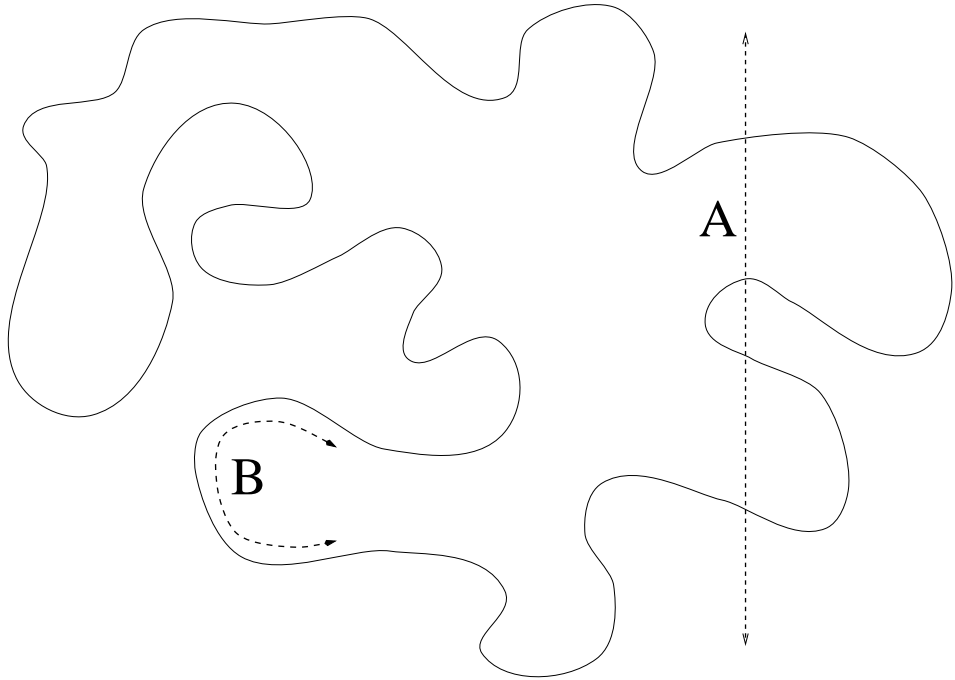,width=5in,angle=270}}
    \caption{A domain in model A.  The width of the OPSF should give an
        indication of how often domains of size $A$ occur. The line $B$ is
        what the dominant length-scale is, as given by the ICAF:
        roughly, the dominant radius of curvature.}
    \label{fig:dominant}
\end{figure}

The dominant length-scale $L_o(t)$ for model A dynamics thus seems to have a
different nature than, for instance, that of model B dynamics, whose OPSF
itself shows a peak at non-zero wavevector.
In the literature on model A and B dynamics one loosely speaks of this dominant
length-scale as an average or typical domain size.  However the concept of
domain size is well defined only when the domains are morphologically
disconnected or, if not, if they have a well-defined width.  As discussed
earlier in this article, model B satisfies this, but for model A only
sufficiently off-critical quenches create domains whose bubble morphology lends
itself to the definition of a ``typical domain size''.
The ICAF for model A shows that a dominant length-scale is present in
the {\em spatial undulations of the interfaces} rather than in the size of the
domains.  The difference is schematized in \fig{dominant}.  We have
referred to this
dominant length as $L_o(t)$, to differentiate it from the degenerate
$L(t)$.  An interesting question this raises is whether model B interfaces 
would exhibit not only the dominant length of domain size, but a dominant 
length of interface undulation as well.

The presence of a dominant {\em undulation} wavelength implies that
model A interfaces may be called ``random'' only approximately
\cite{yeungOS94},  since truly random
interfaces have an uncorrelated ICAF rather than the one found here.  Several
analytical methods developed to derive the scaling function for the OPSF
make use of Gaussian assumptions about the order-parameter field as well as
the randomness of the interfaces \cite{yeungOS94}.
Though for model A these gaussian assumptions appear to work, the
non-gaussian curvature correlations may provide a clue into the
break-down of gaussian assumptions for the more complex model B.

An important difference between the OPSF and the ICSF is that the latter
distinguishes between the two domain morphologies of model A:
bicontinuous for a critical quench, bubble for strongly off-critical quenches.
Indeed, the OPSF is qualitatively the same for both morphologies, whereas for
the bubble morphology the negative curvature autocorrelations in the ICAF
are non-existent:  the peak in the ICSF shifts to $y=0$ for sufficiently
off-critical quenches.  The difference may be due to the absence of phase
information in the OPSF.



\section{Conclusion}

We discussed several features of model A interfacial dynamics via the ICAF
(interface curvature autocorrelation function) $\Gkst$.   Two notable
characteristics of the ICAF were the gaussian form near $s=0$, and the dip
below zero beyond a certain distance.  Though time-dependent lengths can be
defined from the OPSF, and the  existence of a typical and uniquely defined
dynamical length $L_o(t)$ can be infered from the universality of the OPSF, the
peak at zero wavenumber prevents us from uniquely deducing $L_o(t)$ and
identifying its physical meaning.  However the ICAF, through the existence of a
zero, or equivalently the ICSF, through the location of its maximum, clearly
answers both questions.  The scaling length-scale for model A systems is hence
in the {\em interface undulations} of the domains.   This suggests that the
scaling length-scale in model A dynamics is of a different nature than the
domain-size length-scale of model B dynamics, where the order-parameter is
conserved.  It is also clear that the OPSF hides some interesting
characteristics of model A interface dynamics and its domain morphology.  
Experimental
methods for measuring the interface curvature autocorrelations could become
useful in better characterizing the dynamics of some pattern forming systems
where sharp interfaces are present.


\section*{Acknowledgments}

We are grateful to Dr.~Mohamed Laradji and Fran\c{c}ois
L\'eonard for helpful discussions.  We also thank the Natural Sciences and
Engineering Research Council of Canada (NSERC), the Walter C.~Sumner
Foundation and the University of Toronto for partial financial support
during the course of this work.


\end{document}